\documentclass{article}
\usepackage[utf8]{inputenc} 
\usepackage[T1]{fontenc}    
\usepackage{hyperref}       
\usepackage{url}            
\usepackage{booktabs}       
\usepackage{amsfonts}       
\usepackage{nicefrac}       
\usepackage{microtype}      
\usepackage{lipsum}
\usepackage{graphicx}
\graphicspath{ {./images/} }
\usepackage{multirow}
\usepackage{booktabs}
\usepackage{lmodern}
\usepackage{authblk}

\usepackage{amsmath,amssymb,amsfonts}%
\usepackage{mathrsfs}%
\usepackage[title]{appendix}%
\usepackage{xcolor}%
\usepackage{textcomp}%
\usepackage{manyfoot}%
\usepackage{booktabs}%
\usepackage{algorithm}%
\usepackage{algorithmicx}%
\usepackage{algpseudocode}%
\usepackage{listings}%
\usepackage{makecell}
\usepackage{soul,color}
\usepackage[verbose=true,letterpaper]{geometry}
\AtBeginDocument{
  \newgeometry{
    textheight=9in,
    textwidth=6.5in,
    top=1in,
    headheight=14pt,
    headsep=25pt,
    footskip=30pt
  }
}

\title{\textbf{Ab-initio Approach for Constructing Inverse Potentials for Resonant States of $\alpha$ - $^{3}$H and $\alpha$ - $^{3}$He Scattering}}

\author[1]{Ishwar Kant}
\author[1]{Ayushi Awasthi}
\author[1]{Arushi Sharma}
\author[1]{Shikha Awasthi}
\author[1]{O.S.K.S. Sastri\thanks{Email: \texttt{sastri.osks@hpcu.ac.in}}}
\author[2]{M.R. Ganesh Kumar}

\affil[1]{Department of Physics and Astronomical Sciences, Central University of Himachal Pradesh, Dharamshala, 176215, Himachal Pradesh, Bharat (India)}
\affil[2]{Applied Materials India Private Limited, Bengaluru, 560066, Bharat (India)}

\begin{document}
\maketitle
\begin{abstract}
In this paper, the inverse potentials for the resonant $f$ states of $\alpha$-$^{3}\text{H}$ and $\alpha$-$^{3}\text{He}$ are constructed using the phase function method by utilizing an ab-initio approach.\\
\textbf{Central Idea:}- A combination of three piecewise smooth Morse-type functions are chosen over three regions of interaction and are joined smoothly at the intervening boundaries to prepare the reference potential. While the regular Morse function captures the nuclear and Coulomb interactions at short and medium ranges, an inverse Morse function is chosen to obtain the Coulomb barrier that arises because of the long-range Coulomb interaction. This reference potential is representative of a large family of curves consisting of eight distinct model parameters and two intermediate points that define the boundaries that exist between the three regions. \\
\textbf{Methodology:} The phase equation is solved using the Runge-Kutta 5$^{th}$ order method for the input reference potential to obtain the scattering phase shifts at various center of mass energies. The model parameters are then adjusted using the genetic algorithm in an iterative fashion to minimize the mean square error between the simulated and expected phase shift values.\\
\textbf{Results:} Our approach successfully constructed the inverse potentials for the resonant $f$ states of the $\alpha$-$^{3}\text{H}$ and $\alpha$-$^{3}\text{He}$  systems, achieving convergence with a minimized mean square error. The resonance energies and widths for the $\alpha$-$^{3}\text{H}$ system for the $f_{5/2}^{-}$ and $f_{7/2}^{-}$ states are determined to be [4.19 (4.14), 1.225 (0.918)] MeV and [2.20 (2.18), 0.099 (0.069)] MeV, respectively. For the $f_{5/2}^{-}$ and $f_{7/2}^{-}$ states of the $\alpha$-$^{3}\text{He}$ system, the resonance energies and widths are [5.03 (5.14), 1.6 (1.2)] MeV and [2.99 (2.98), 0.182 (0.175)] MeV, respectively. \\ 
\textbf{Conclusion:} Our ab-initio approach to solve the phase equation utilizing a combination of smoothly joined Morse functions effectively captures both short-range nuclear and long-range Coulomb interactions, providing an accurate model for nuclear scattering involving charged particles.
\end{abstract}


\section{Introduction}\label{sec1}

The $\alpha$-$^{3}H$ and $\alpha$-$^{3}He$ reactions are crucial in both nuclear and astrophysical studies, particularly in understanding light nuclei. Nucleosynthesis processes in the universe have yielded a plethora of chemical elements, with heavier ones resulting from nuclear reactions and lighter ones, such as hydrogen, helium, and lithium, originating from the Big Bang. The primordial abundance of \(^{7}Li\) is shaped by the \(^{3}H\)(\(\alpha, \gamma\))\(^{7}Li\) and the abundance of \(^{7}Be\) is shaped by the \(^{3}He\)(\(\alpha, \gamma\))\(^{7}Be\) radiative-capture process. Both processes are essential for solar hydrogen burning and neutrino production.\\
Since 1960's\cite{Christy}, the $\alpha$-$^{3}He$ reaction is an interesting problem and has been extensively studied by both experimentalists and theoreticians due to its significance in nuclear physics and astrophysics\cite{Dario}. In the solar hydrogen burning pp-chains also, these reactions are the beginning of the pp-II and pp-III reaction branches. Acquiring accurate data on this reaction has widened significantly in recent years as it has lately become the second most significant source of uncertainty among nuclear inputs in computing the high-energy solar neutrino flux. These are also an important problem in halo effective field theory (hEFT) \cite{Hammer}. At around 23 keV, the most astrophysical significant energy\cite{Nara}, theoretical extrapolations, such as those using microscopic models \cite{Kajino, Kenneth}, of data collected at higher energies are required since it is practically impossible to perform investigations at such low energies in laboratories. Additionally, the cross-section for this reaction can be easily measured at energies higher than 110 keV. However, at low energies near 20 keV, measuring the cross section is challenging due to the Coulomb barrier, which causes exponential suppression at low energies\cite{Haxton}. \\
Nuclear data input plays a crucial role in determining the accurate synthesis of light elements in astrophysical environments \cite{Krauss}. Although less extensive work has been done on the $\alpha$-$^{3}H$ reaction, recent studies have shown growing interest in exploring the scattering phase shifts, scattering cross sections, and the astrophysical S-factor of this reaction. In 1993, Michael S. Smith $et.al$ \cite{Smith} did a comprehensive evaluation of 12 nuclear reactions by conducting detailed study on their reaction rates and uncertainties for the production of light nuclei and found $\alpha$-$^{3}H$ reaction to be the most uncertain of all the 12 reactions. After their study in 1994, C. R. Brune $et.al.$ \cite{Brune} performed an experiment on $\alpha$-$^{3}H$ at low energies where they obtained the absolute cross sections for center of mass energies ranging between 0.05 MeV and 1.2 MeV with an uncertainty $\approx$ 6 $\%$. Recently, in 2016, Jérémy Dohet-Eraly\cite{Dohet} $et.~al.$ calculated astrophysical S-factor for both $\alpha$-$^{3}H$ and $\alpha$-$^{3}He$ reactions within $ab~initio$ method i.e no core shell model with continuum (NCSMC) using a renormalized chiral nucleon–nucleon interaction.\\
From a nuclear physics standpoint, the interaction of odd-A nuclei, such as \(^{3}H\) and \(^{3}He\), with \(\alpha\) particles leads to compound nuclei like \(^{7}Li\) and \(^{7}Be\), which are also odd-A magic nuclei. These reactions typically involve the scattering of spin-\(\frac{1}{2}\) particles with spin-0 particles in their entrance channels. The importance of these reactions in nuclear physics and astrophysics makes them crucial for the study of light nuclei as well.\\
 In 1958, Philip D. Miller and his team\cite{Miller} conducted experiments on the scattering of \( ^3{He}\) from \( ^4{He}\), analyzing excitation energies in \( ^7{Be}\) between 3.28 and 4.73 MeV. They also plotted and interpreted the phase shifts for $s(\ell=0)$, $p(\ell=1)$, $d(\ell=2)$, and $f-waves(\ell=3)$ at bombarding energies below 6 MeV. In 1962, T. A. Tombrello and P. D. Parker \cite{Tombrello} explored the excited states in \( ^7 \text{Be} \) by scattering doubly charged \( ^3 \text{He}\) ions from the helium gas, focusing on excitation energies between 3.9 and 8.4 MeV. They derived phase shifts for $s$, $p$, $d$, and $f$ waves from angular distributions and smoothed excitation functions. A. C. L. Barnard $et.~al.$ \cite{Barnard} in 1963, measured the cross section for elastic scattering of $^3He$ by $^4He$ for incident energies in the range 2.5-5.7 MeV and measured excitation functions at center-of-mass angles starting from 54 degrees to 140 degrees.  R. J. Spiger and T. A. Tombrello \cite{Spiger} in 1967, measured differential elastic scattering cross section of $\alpha$-$^{3}H$ and $\alpha$-$^{3}He$ scattering for projectile energies of 4 to 18 MeV and 5 to 18 MeV respectively and suggested $7/2^-$ and $5/2-$ assignments in \(^{7}Li\) and \(^{7}Be\), by including new energy levels at 9.7 MeV and 9.3 MeV.\\ 
 In 1968 M. Ivanovich and his team\cite{Ivanovich} studied the elastic scattering of protons, deuterons, \( ^3{He}\), and \( ^4{He} \) particles of tritium, covering incident energies ranging from 3 to 11 MeV. L.L. Chopovsky investigated scattering length for both $\alpha$-$^3H$ and $\alpha$-$^3He$ systems in 1989, by utilising resonating group method (RGM) and considering the limiting case when the relative motion energy of colliding clusters is equal to zero \cite{Chopovsky}.\\
Recently in 2009, Peter Mohr \cite{Mohr} studied cross section for $\alpha$-$^3He$ reaction at low energies using a simple two-body interaction within direct capture (DC) model in combination with a double-folding potential\cite{Mohr}. In 2017, Tamás Szücs $et.~al.$ obtained cross sections for centre of mass energies between 2.5-4.4 MeV. Matteo Vorabbi\cite{Vorabbi} $et.~al.$ also studied \(^{7}Li\) and \(^{7}Be\) nuclei within NCSMC and analyzed all the binary mass partitions involved in the formation of these systems. Lately, Laha group showed interest in $\alpha$-$^3H$ and $\alpha$-$^3He$ systems in 2018, by considering Hulth{\'e}n potential as the interaction potential and obtained scattering phase shifts by utilising phase function method (PFM). Phase Function Method (PFM) is a vital technique in quantum physics, indispensable for calculating phase shifts that describe particle behavior within a system. Initially introduced by Morse and Allis \cite{Morse} for the determination of specific phase shifts, PFM has been refined and expanded by contributions from Drukarev \cite{Drukarev}, Bergmann \cite{Bergmann}, Kynch \cite{Kynch}, Calogero \cite{Calogero,Calogero_Old} and Babikov\cite{Babikov}.\\
A major advantage of the Phase Function Method (PFM) is its efficiency, as it only requires the interaction potential $V(r)$ for phase shift calculations, eliminating the need for wave function computations, unlike other theoretical methods like R-matrix, S-matrix, Jost function, or complex-scaling, which necessitate wave function calculations. PFM accurately determines the scattering phase shifts by integrating a first-order nonlinear differential equation numerically. Thus to implement this method the need of experimental data is crucial. We have already used this method successfully to obtain scattering phase shifts for \( np \), \( \alpha-\alpha \) \cite{alpha, awasthi2024comparative}, and \( nd \) scattering \cite{sawasthi}, utilizing Morse potential, Gaussian potential, and Malfliet-Tjon potential. Our group \cite{IJP} has also done calculations for $\alpha$-$^3H$ and $\alpha$-$^3He$ systems, using Malfliet-Tjon potential as the interaction potential and Hulth{\'e}n as the screened Coulomb potential within the low energy region by employing PFM.\\
Since, $\alpha$-$^{3}H$ and $\alpha$-$^{3}He$ systems are charged systems, therefore interaction between both the interacting particles is a combination of short range nuclear part and the long range Coulomb part. In this paper, we have focused  on constructing inverse potential for $\alpha$-$^{3}H$ and $\alpha$-$^{3}He$ elastic scattering below 11 MeV utilising three piece-wise Morse functions, joined smoothly, to describe the distinct regions of $\alpha$-$^{3}H$ and $\alpha$-$^{3}He$ interactions. The experimental scattering phase shifts have been taken from R.J. Spiger and T. A. Tombrello \cite{Spiger}, WR Boykin \cite{Boykin} and DM Hardy \cite{Hardy}.

\section{Methodology:}
\subsection{Phase Function Method:}

In phase function method (PFM), the second-order linear homogeneous Schrödinger equation is transformed into a first-order nonlinear Riccati equation\cite{Calogero}. This reduces the scattering problem from obtaining information through wave-function to one of directly finding the scattering phase shifts. Knowing the phase function allows for the complete determination of the wave function at different bound and scattering energies and also calculate other observable quantities such as partial scattering amplitudes, whose poles indicate bound state energies. Initially developed for spherically symmetrical potentials, the PFM has since been extended to more complex scenarios, including non-central forces, multichannel scattering, and relativistic equations. Thus, it can be used to formulate and solve any quantum-mechanical problem involving scattering or bound states.
Partial wave analysis is used to derive the phase shift (\(\delta\)) for different orbital angular momentum states of the incoming and outgoing waves, which are modified due to the interaction between the projectile and the target nucleons.
The Schrödinger wave equation for a spinless particle with energy \( E \) and orbital angular momentum \( \ell \) that undergoes scattering is expressed as
\begin{equation}
\frac{\hbar^2}{2\mu} \left[\frac{d^2}{dr^2}+\left(k^2-\frac{\ell(\ell+1)}{r^2}\right)\right]u_{\ell}(k,r) = V(r)u_{\ell}(k,r)
\label{Scheq}
\end{equation}
This second-order differential equation can be transformed into a first-order non-homogeneous differential equation of the Riccati type \cite{Calogero, Babikov}, given by
\begin{equation}
\delta_{\ell}'(k,r) = -\frac{V(r)}{k \left(\frac{\hbar^2}{2\mu}\right)} \left[\cos(\delta_\ell(k,r)) \hat{j}_{\ell}(kr) - \sin(\delta_\ell(k,r)) \hat{\eta}_{\ell}(kr)\right]^2
\label{PFMeqn}
\end{equation}
Here, the prime denotes differentiation of the phase shift with respect to distance. The Riccati-Hankel function of the first kind relates to the Ricatti-Bessel function (\(\hat{j}_{\ell}(kr)\)) and the Riccati-Neumann function (\(\hat{\eta}_{\ell}(kr)\)) as \(\hat{h}_{\ell}(r) = -\hat{\eta}_{\ell}(r) + \textit{i} \hat{j}_{\ell}(r)\). The Ricatti-Bessel and Riccati-Neumann functions can be derived using the recurrence relations:
\begin{equation}
\hat{j}_{\ell+1}(kr) = \frac{2 \ell + 1}{kr} \hat{j}_{\ell}(kr) - \hat{j}_{\ell-1}(kr)
\end{equation}
\begin{equation}
\hat{\eta}_{\ell+1}(kr) = \frac{2 \ell + 1}{kr} \hat{\eta}_{\ell}(kr) - \hat{\eta}_{\ell-1}(kr)
\end{equation}
In this work, we are interested in solving the phase equation for $\ell = 3$ channel, given by
\begin{eqnarray}
\delta_{3}'(k,r) = -\frac{V(r)}{k \left(\frac{\hbar^2}{2\mu}\right)} \left[\frac{(kr)^3 \cos(\delta_3(k,r) + kr) - 6(kr)^2 \sin(\delta_3(k,r) + kr)}{(kr)^3} - \right. \nonumber \\
\left. \frac{15(kr) \cos(\delta_3(k,r) + kr) + 15 \sin(\delta_3(k,r) + kr)}{(kr)^3}\right]^2
\label{PFMeqn3}
\end{eqnarray}

The function \(\delta_{3}(k,r)\) in Equation \ref{PFMeqn3} is known as the phase function, and its value at any point \( r = R \) provides the phase shift corresponding to the interaction potential \( V(r) \) at that point. Once, the interaction ceases, the phase shift value becomes constant.

Equation \ref{PFMeqn3} is a non-linear differential equation that can be solved from the origin to the asymptotic region to obtain scattering phase shifts for various energies using the initial condition \(\delta_{\ell}(0) = 0\). The equation is numerically solved using the Runge-Kutta fifth-order (RK-5) method. 
In this paper, our aim is to construct the model interaction potential for the chosen scattering system. To achieve this, we utilize an ab initio framework that starts from first principles, allowing us to accurately construct the interaction between particles in the system. This approach considers a family of possible smooth curves as reference input to Equation \ref{PFMeqn3} for arriving at the correct optimized potential that gives rise to the expected scattering phase shifts while ensuring that the modeled interaction is physically meaningful. 

\subsection{Ab-initio Approach for Constructing Inverse Potential}
In scattering theory, the interaction potential between particles can be typically divided into three main regions, each characterized by distinct physical behaviors that are critical to accurately model for reliable scattering predictions. The first region is that corresponding to very short distances, which is dominated by a strong repulsive core due to the Pauli exclusion principle and the short-range nature of the nuclear force, which prevents projectiles from coming too close to the target. This repulsion is crucial for ensuring stability in nuclear systems, as it balances the attractive nuclear force at slightly larger separations. The second, or intermediate, region, corresponds to the range where the attractive nuclear forces dominates, forming a potential well that is responsible for binding and reveals nuclear structure. Beyond a certain distance, typically r$\geq$ 2 to 3~fm, the interaction potential gradually decays, transitioning into a long-range tail where the effects of nuclear forces diminish, and in the case of charged particles, Coulomb interactions become significant. An effective potential model should thus account for all three regions—short-range repulsion, intermediate attraction, and long-range decay—to accurately describe the dynamics of nuclear scattering.

In this study, we constructed the reference potential using an ab initio approach, meaning we avoided assuming any specific predefined functional form for the interaction based on physical considrations. This approach allows the flexibility needed to explore the genuine physical nature of the interaction without bias. Several mathematical functions, including widely used forms such as the Malfliet-Tjon potential, were initially tested to see if they could serve as suitable models. However, we found that these potentials fell short in accurately capturing the inverse potential’s detailed structure, particularly in reproducing the correct behavior across the different interaction regions. Gaussian functions, commonly utilized for their simplicity, also proved inadequate due to their inability to capture the strong repulsive core at short distances effectively. This emphasizes the importance of selecting mathematical functions capable of accurately representing the distinct characteristics of the interaction, such as repulsive short-range forces, binding in the intermediate range, and the long-range tail that also captures the Coulomb barrier for charged systems. 

Consequently, we adopted Morse-type functions to construct the potential\cite{Selg}, as it is the combination of exponential terms which allows them to effectively capture the complex structure of the interaction potential, including the strong short-range repulsion and gradual transition to long-range behavior. The Morse potential is particularly advantageous as it can be adapted to the different interaction regions through parameter tuning, making it suitable for modeling nuclear interactions that vary with distance.

Previously, we utilized a configuration of two and three piecewise, smoothly joined Morse functions as a reference potential, and effectively described the $\alpha$-$\alpha$ \cite{alpha_alpha} and neutron-proton\cite{np} systems. In this paper, we have chosen three Morse functions as reference function to construct the inverse potentials with the first two terms, $V_{R}$, $V_{NC}$, to represent the short-range nuclear forces including the Coulomb interaction, and the third term $V_{CL}$, used in an inverted form, to account for the long-range Coulomb interaction, particularly crucial for accurately describing scattering behavior in charged systems.

Thus, the reference potential $V(r)$ is defined as:

\begin{eqnarray*}
V(r) =
\begin{cases} 
V_{R}(r) & \text{if } r \leq x_1 \\
V_{NC}(r) & \text{if } x_1 < r \leq x_2 \\
V_{CL}(r) & \text{if } r > x_2 
\end{cases}
\end{eqnarray*}
To ensure the smoothness of potential at the boundary points $x_1$ and $x_2$ between the three functions,the functions and their derivatives need to be continuous at $ x_1 $ and $ x_2 $. That is:

 \begin{eqnarray*}
V_{R}(r) \big|_{x_1} = V_{NC}(r) \big|_{x_1}, \quad V_{NC}(r) \big|_{x_2} = V_{CL}(r) \big|_{x_2}
 \end{eqnarray*}

\begin{eqnarray*}
\left. \frac{dV_{R}(r)}{dr} \right|_{x_1} = \left. \frac{dV_{NC}(r)}{dr} \right|_{x_1}, \quad \left. \frac{dV_{NC}(r)}{dr} \right|_{x_2} = \left. \frac{dV_{CL}(r)}{dr} \right|_{x_2}
\end{eqnarray*}

where, $V_{R}$, $V_{NC}$ and $V_{CL}$ accounts for short range nuclear interaction, short range Nuclear-Coulomb interaction and long range coulomb interaction potential respectively.


\subsection*{Mathematical Formulations}
The three Morse components of the reference function $V_{R}$, $V_{NC}$ and $V_{CL}$ are defined as:
\begin{eqnarray*}
V_R = V_0 + D_0 \left( e^{-2 \alpha_0 (r - r_0)} - 2 e^{-\alpha_0 (r - r_0)} \right)
\end{eqnarray*}
\begin{eqnarray*}
V_{NC} = V_1 + D_1 \left( e^{-2 \alpha_1 (r - r_1)} - 2 e^{-\alpha_1 (r - r_1)} \right)
\end{eqnarray*}
\begin{eqnarray*}
V_{CL} = V_2 - D_2 \left( e^{-2 \alpha_2 (r - r_2)} - 2 e^{-\alpha_2 (r - r_2)} \right)
\end{eqnarray*}
As each function $V_{R}$, $V_{NC}$ and $V_{CL}$ have 4 parameters, and $x_1$ and $x_2$ must also be varied, we have a reference potential with a total of 14 parameters. Applying the 4 boundary conditions determines the parameters $D_1$, $D_2$, $V_0$, and $V_1$, reducing the total number of model parameters to 10. Therefore, the reference model consists of a family of curves defined by 8 parameters, with boundary points $x_1$ and $x_2$.
\( D_1 \) and \( D_2 \) are obtained utilizing the continuity conditions at the boundaries as:
\begin{eqnarray}
D_1 = \frac{\alpha_0 D_0 g_0}{\alpha_1 g_1}
\end{eqnarray}
\begin{eqnarray}
D_2 = -\frac{\alpha_1 D_1 h_1}{\alpha_2 h_2}
\end{eqnarray}
where the factors
\( g_0 \), \( g_1 \) and \( h_0 \), \( h_1 \) are defined as
\begin{eqnarray}
g_0 = e^{-2 \alpha_0 (x_1 - r_0)} - e^{-\alpha_0 (x_1 - r_0)},
\qquad
g_1 = e^{-2 \alpha_1 (x_1 - r_1)} - e^{-\alpha_1 (x_1 - r_1)}
\end{eqnarray}
\begin{eqnarray}
h_1 = e^{-2 \alpha_1 (x_2 - r_1)} - e^{-\alpha_1 (x_2 - r_1)},
\qquad
h_2 = e^{-2 \alpha_2 (x_2 - r_2)} - e^{-\alpha_2 (x_2 - r_2)}
\end{eqnarray}
Furthermore, $V_0$ and $V_1$ are shift parameters that are expressed as functions of $V_2$, where $V_2$ serves as the primary optimizable parameter, as follows:
\begin{eqnarray}
V_1 = V_2 - {D_2}{j_2} -{D_1}{j_1}
\end{eqnarray}
\begin{eqnarray}
V_0 = V_1 + D_1 f_1 - D_0 f_0
\end{eqnarray}
where the factors $f_0$, $f_1$ and $j_1$, $j_2$ are defined as
\begin{eqnarray}
f_0 = e^{-2 \alpha_0 (x_1 - r_1)} - 2 e^{-\alpha_0 (x_1 - r_1)},
\qquad
f_1 = e^{-2 \alpha_1 (x_1 - r_2)} - 2 e^{-\alpha_1 (x_1 - r_2)}
\end{eqnarray}
\begin{eqnarray}
j_1 = e^{-2 \alpha_1 (x_2 - r_1)} - 2 e^{-\alpha_1 (x_2 - r_1)},
\qquad
j_2 = e^{-2 \alpha_2 (x_2 - r_2)} - 2 e^{-\alpha_2 (x_2 - r_2)}
\end{eqnarray}
These relationships help maintain consistency and reduce the number of free parameters in the model. The functions \( f_0 \), \( f_1 \), \( g_0 \), \( g_1 \), \( j_1 \), \( j_2 \), \( h_1 \), and \( h_2 \) are combinations of exponential terms. By defining these equations and relationships, we can construct V(r) as the interaction potential within the system, allowing us to solve the phase equation and calculate the scattering phase shifts.



\subsection{Optimisation Technique:}
To construct the inverse potential from the experimentally available data, from the designed reference system that constitutes a family of smooth curves, we must optimize the model parameters. For this purpose, we employ a machine-learning-based genetic algorithm, which iteratively evolves parameter sets to minimize the error between the predicted and target values by minimizing a cost function. This method effectively searches for optimal solutions by simulating the process of natural selection, and refining the potential model that accurately represents the underlying physical interactions.

\subsubsection*{Procedure to construct the Inverse Potential:}
\begin{itemize}
\item \textbf{Input Data:} The experimental lab energies $E_{lab}$ of the projectile and the corresponding scattering phase shifts $\delta(E)$ that are obtained from the experimental cross-sections are considered.  Based on the masses of the projectile and the target, the reduced mass of the system is determined and the center of mass energies $E_{CM}$ for the chosen lab energies are calculated. A plot of phase shifts $\delta(E)$ as a function of $E_{CM}$ would tell us about the possible resonances based on the nature of variation of the slope. This would help in determining the region of interest and careful sampling of the experimental energies in order to curate the data to be considered for optimization procedure. 
\item \textbf{Defining the Reference Potential:} Specify a range of values for each model parameter. This gives a broad set of curves to begin with. From this set, randomly select initial values for each parameter to construct the ab-initio reference potential. 
\item \textbf{Numerical Solution of the Phase Equation:} To compute the simulated scattering phase shifts (SPS), denoted as $\delta_i^{sim}$, one solves the phase equation numerically using the fifth-order Runge-Kutta (RK-5) method. This solution uses the initialized reference potential as input, providing a simulated scattering phase shifts that can be compared with the expected data obtained from experimental cross-sections.
\item \textbf{Calculating Cost Function:}
To assess the accuracy of the simulated scattering phase shifts, we calculate the mean squared error (MSE) between the simulated SPS ($\delta_i^{sim}$) and the experimental data ($\delta_i^{exp}$). The MSE is computed as:
\begin{equation}
MSE= \frac{1}{N}\sum_{i=1}^{N}{(\delta^{exp}_i-\delta^{sim}_i)^2}
\end{equation}
where N is the total number of data points. The MSE quantifies how closely the simulated data matches the experimental results, with lower values indicating a better fit.
\item \textbf{Optimization via Genetic Algorithm:} Using the MSE as the cost function, the optimization process is carried out using a genetic algorithm (GA). In each iteration of this algorithm, a population of candidate solutions (sets of model parameters) is generated, and their fitness is evaluated based on the MSE. The solutions with the lowest MSE are selected for reproduction, where genetic operators such as crossover and mutation create new offspring. This process iterates over multiple generations, refining the model parameters to minimize the cost function and improve the fit to the experimental data.
\item \textbf{Convergence:} The optimization procedure continues until the MSE stabilizes, indicating that the algorithm has converged and the model parameters have reached their optimal values. At this point, the construction of the inverse potential is complete, providing the most accurate representation of the scattering system based on the given experimental data. 
\item \textbf{Testing and Validation}: One can start with different initial seeds to ensure reproducibility. One must also consider different ranges for the model parameters so that a possible curve with a smaller MSE is not missed out. This is also required to ensure that constructed inverse potential is physically meaningful.

\end{itemize}

\section{Results and Discussions}
\subsection{Database:}
The scattering phase shift (SPS) data for various $\ell$-channels has been taken from R.J. Spiger and T. A. Tombrello \cite{Spiger}, WR Boykin \cite{Boykin} and DM Hardy \cite{Hardy}. Even though the SPS data is available for various S, P and F states, only the resonant F-states have significant values and correspond to observed resonances. Hence, in this paper, we focus on the resonance states of the $\alpha$-$^{3}\text{H}$ and $\alpha$-$^{3}\text{He}$ systems, specifically the $\ell=3$ state with negative parity. As detailed in our previous study \cite{IJP}, we categorized the energies up to 11 MeV into three distinct regions: 0-5 MeV, 5-6 MeV, and 6-11 MeV for the $f_{5/2}^-$ state and 0-4 MeV, 4-6 MeV, and 6-11 MeV for the $f_{7/2}^-$ states. This classification is based on a comprehensive analysis of data trends. The final datasets used for calculations are highlighted in the Table \ref{table1} and \ref{table2} given in Appendix. We observed that the phase shifts remain relatively constant in the first and third regions. In contrast, the second region exhibits a significant variation in $d \delta(E)/dE$, which is indicative of the resonance behaviour in these states.
\subsection{Constructing the Inverse Potentials:}
\begin{figure}[htbp]
    \centering
    \includegraphics[scale=0.7]{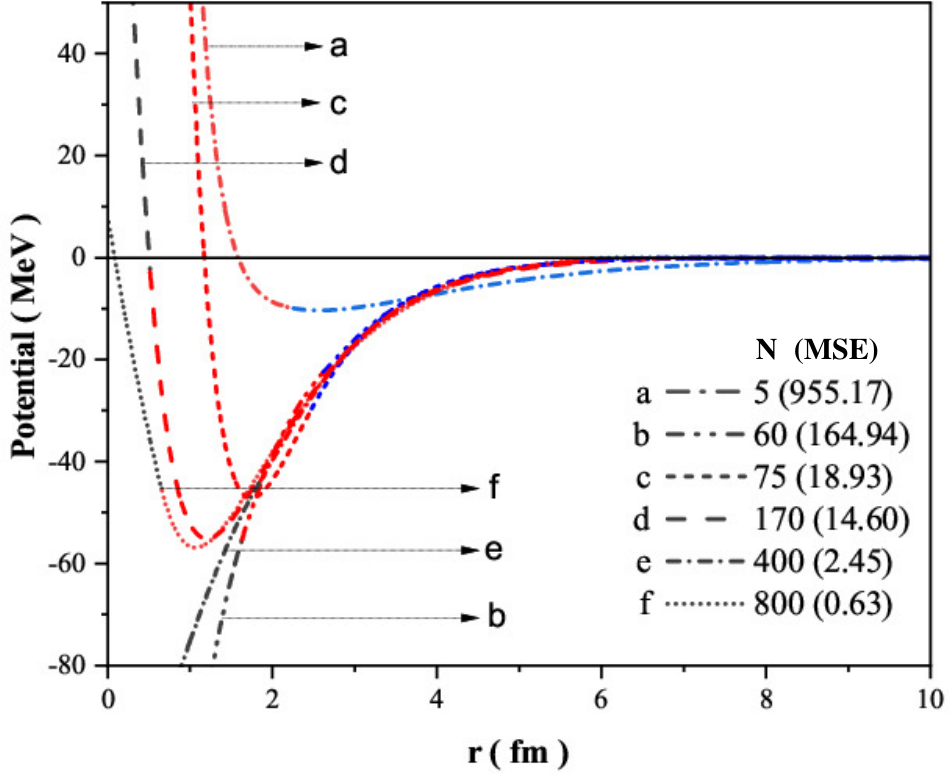}
    \caption{Visualization of different potential graphs obtained during optimization. As the number of iterations(N) increases, the MSE is seen to be decreasing. }
    \label{diff_col}
\end{figure}

Using the chosen datasets, we have optimized the model parameters in the reference function and constructed the inverse scattering potentials for the resonant states of $\alpha$-$^{3}\text{H}$ and $\alpha$-$^{3}\text{He}$. To achieve this, we generated a large parameter space by defining specific bounds for each parameter, resulting in a wide range of possible curves. The first set of model parameters are generated randomly from these interval bounds specified for each parameter, by starting with a random seed. Then, the curve with the least mean squared error (MSE) at the end of certain number of iterations has been plotted, as shown in Fig.\ref{diff_col}, to illustrate the process of optimization, as the genetic algorithm evolves through natural selection based on crossover and mutations. 
Initially, with fewer iterations, the MSE can be seen to be quite large, indicating that the constructed inverse scattering potential was not accurate. However, as the number of iterations increased, the MSE decreased, thus resulting in required inverse scattering potential. Ultimately, the optimal potential is identified as the one that best matches the expected phase shifts. 
\subsection{\texorpdfstring{Inverse Potentials for $\alpha$-$^{3}\text{H}$ and $\alpha$-$^{3}\text{He}$ interactions:}{Inverse Potentials for alpha-3H and alpha-3He interactions:}}
The optimized model parameters obtained for the $\ell=3$ states of $\alpha$-$^{3}\text{H}$ and $\alpha$-$^{3}\text{He}$ are given in Table \ref{optimized}. During optimization, we observed that the value of the parameter $V_2$ tends to approach zero or is on the order of $10^{-8}$ and is not shown explicitly.

\begin{table}[ht!]
\caption{Optimized model parameters for $\alpha$-$^{3}H$ and $\alpha$-$^{3}He$ interactions.}
\renewcommand{\arraystretch}{1.3} 

\begin{tabular}{ccccccccccc} 
\hline
\Xhline{1pt}
\textbf{System}	&\textbf{State}& \textbf{$\alpha_0$}	&	\textbf{$\alpha_1$}	&	\textbf{$\alpha_2$}&	\textbf{$r_0$}&	\textbf{$r_1$}&	\textbf{$r_2$}&	\textbf{$x_1$}&	\textbf{$x_2$}&	\textbf{$D_0$}	\\
\hline
\multirow{2}{*}{\textbf{$\alpha$-$^{3}$H}} & $5/2^-$ & 1.004 & 1.233 & 2.426 & 0.937 & 1.647 & 0.494 & 2.135 & 6.759 & 68.628 \\ 
                            & $7/2^-$ & 0.650 & 1.299 & 1.791 & 0.874 & 0.961 & 4.729 & 1.122 & 6.291 & 257.394 \\ 
\hline

\multirow{2}{*}{\textbf{$\alpha$-$^{3}$He}} & $5/2^-$ & 1.051 & 0.939 & 1.941 & 1.279 & 0.521 & 2.905 & 2.827 & 3.904 & 62.199 \\ 
                             & $7/2^-$ & 1.169& 0.712& 1.696& 1.037& 0.380& 1.579& 1.705& 2.534& 94.680\\

\Xhline{1pt}
\hline

\end{tabular}
\label{optimized}
\end{table}

\begin{figure}[htbp]
\begin{tabular}{ccc}
\includegraphics[scale=0.37]{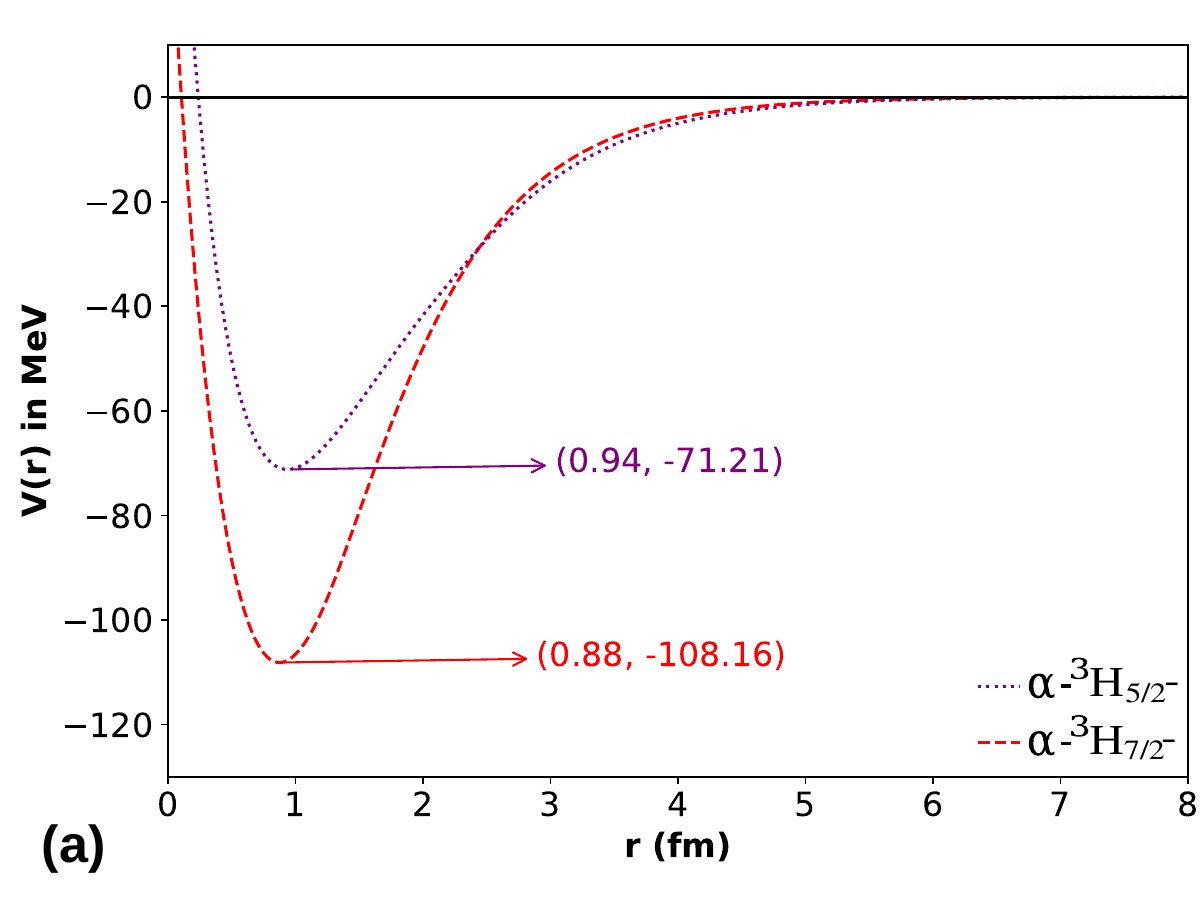}& \includegraphics[scale=0.37]{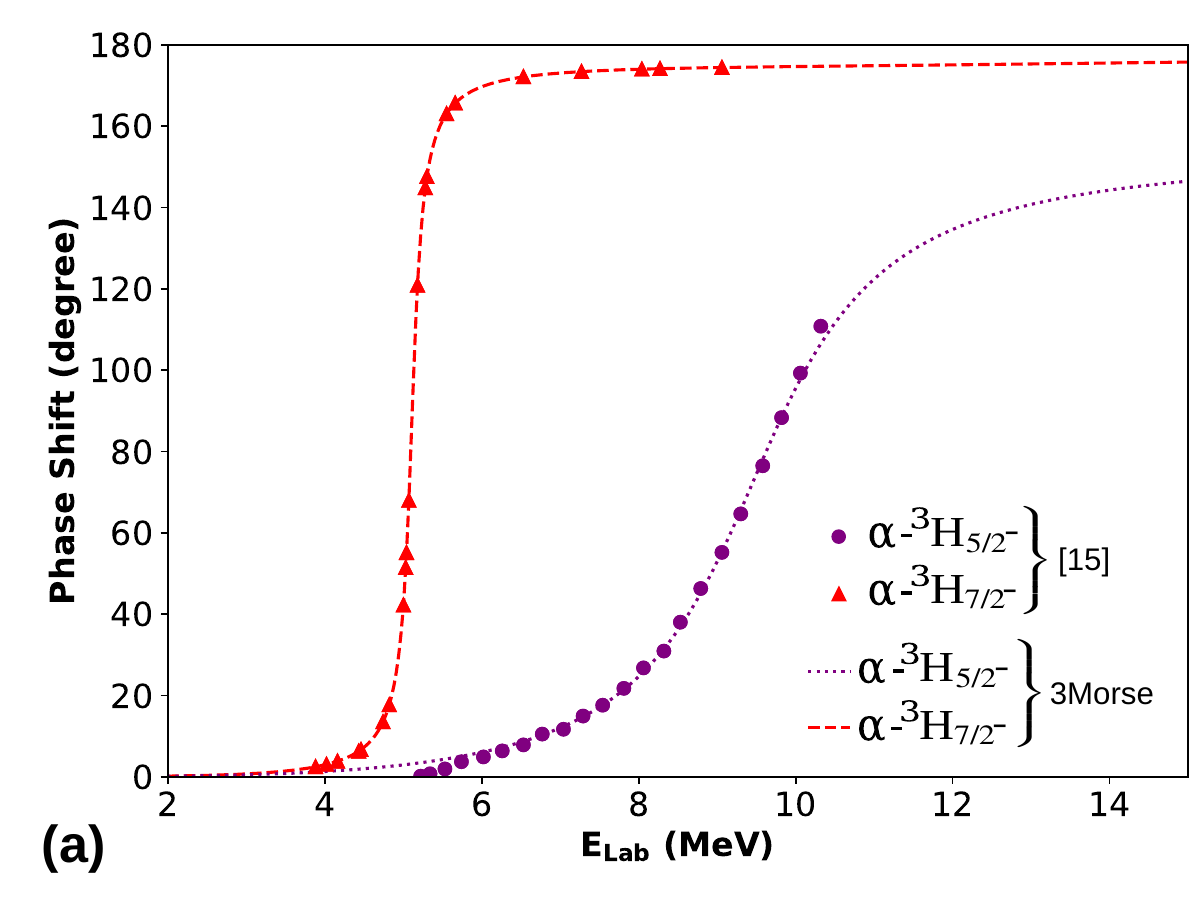}&
\end{tabular}
 \caption{(a) The interaction potential without the centrifugal term and (b) the associated scattering phase shifts for the $5/2^-$ and $7/2^-$ resonant states in $\alpha$-$^{3}$H scattering.}
\label{ptn_ps1}
\end{figure}
\begin{figure}[htbp]
\begin{tabular}{ccc}
\includegraphics[scale=0.37]{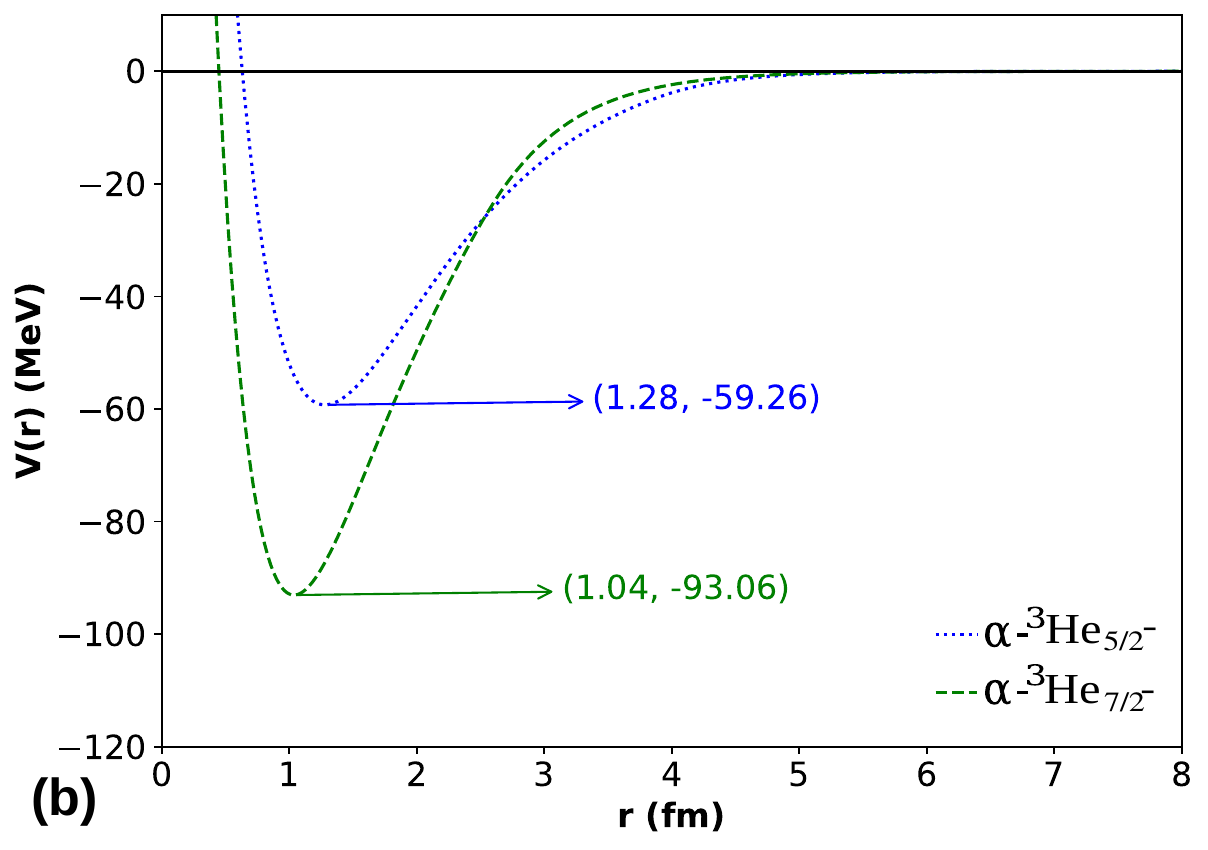}& \includegraphics[scale=0.37]{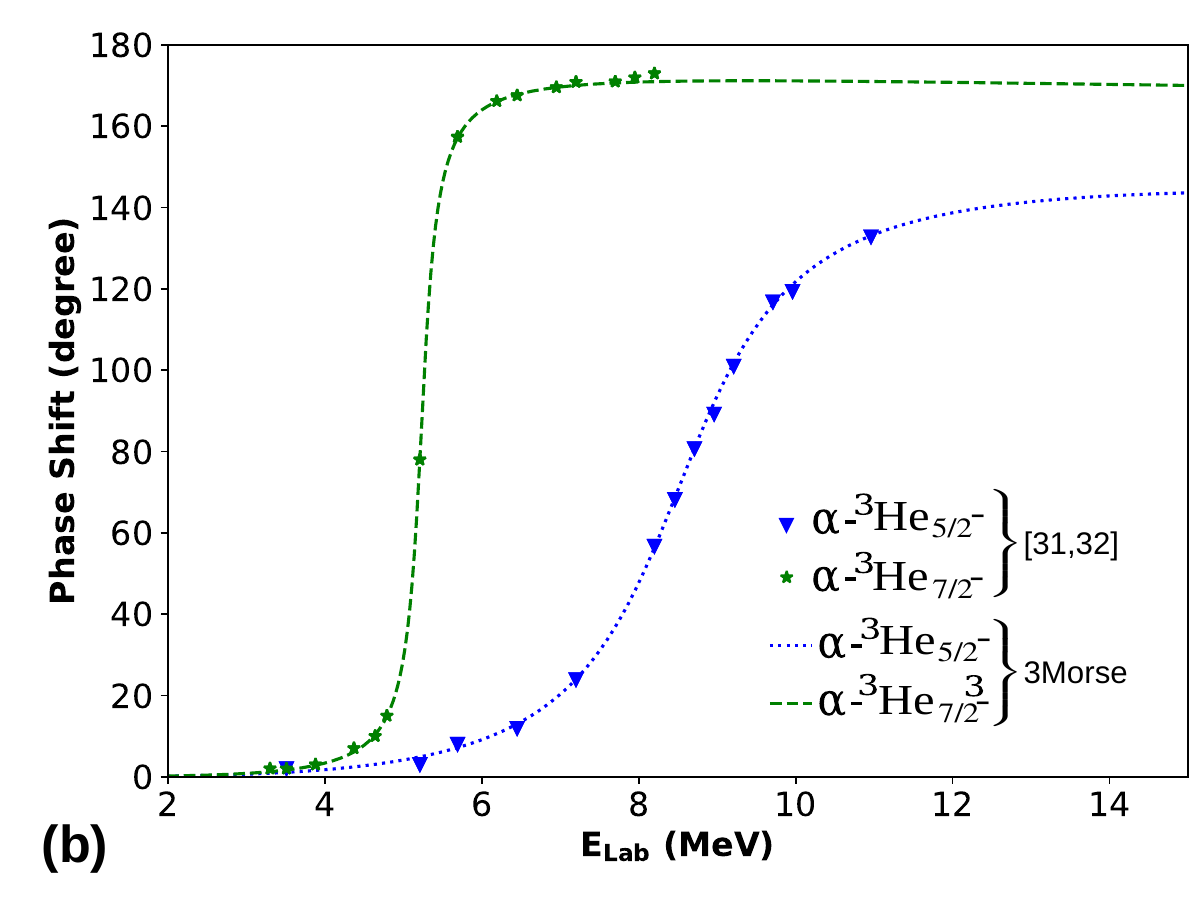}&
\end{tabular}
 \caption{(a) The interaction potential without the centrifugal term and (b) the associated scattering phase shifts for the $5/2^-$ and $7/2^-$ resonant states in $\alpha$-$^{3}$He scattering.}
\label{ptn_ps2}
\end{figure}

\begin{table}[ht]
\centering

\caption{Interaction parameter for $\alpha$-$^{3}$H and $\alpha$-$^{3}$He resonant $f$-wave.}
\renewcommand{\arraystretch}{1.2}
\begin{tabular}{cc c c c ccc}
\hline
\Xhline{1pt}

\textbf{System} & \textbf{State} & \(\mathbf{V_{d}}\) & \(\mathbf{r_{d}}\) & \(\mathbf{V_{d}}^{cent}\) & \(\mathbf{r_{d}}^{cent}\) & \(\mathbf{V_{CB}}^{cent}\) & \(\mathbf{r_{CB}}^{cent}\) \\
 & & \textbf{(in MeV)} & \textbf{(in fm)} & \textbf{(in MeV)} & \textbf{(in fm)} & \textbf{(in MeV)} & \textbf{(in fm)} \\
\hline
\multirow{2}{*}{\textbf{$\alpha$-$^{3}$H}} & ${5/2^-}$ & -71.21 & 0.94 & -5.63 & 2.13 & 4.53 & 4.59 \\  
 & ${7/2^-}$ & -108.16 & 0.88 & -15.52 & 1.63 & 5.19 & 4.25 \\ 
\hline
\multirow{2}{*}{\textbf{$\alpha$-$^{3}$He}} & ${5/2^-}$ & -59.26 & 1.28 & -5.42 & 2.09 & 5.72 & 4.40 \\  
 & ${7/2^-}$ & -93.06& 1.04& -15.06& 1.71& 6.79& 3.85\\ 
\hline
\Xhline{1pt}
\end{tabular}
\label{interaction}
\end{table}


The inverse potential obtained with these optimized parameters for $\alpha$-$^{3}\text{H}$ and $\alpha$-$^{3}\text{He}$ are plotted in Fig.\ref{ptn_ps1}(a) and Fig.\ref{ptn_ps2}(a) respectively. The corresponding phase shift obtained along MSE expected are shown in Fig.\ref{ptn_ps1}(b) and \ref{ptn_ps2}(b) respectively.

From the constructed potential one can deduce the interaction parameters such as potential depth, the equilibrium distance at which the attraction is strongest, as follows:
\begin{enumerate}
    \item For the $5/2^{-}$ state of $\alpha$-$^{3}\text{H}$ and $\alpha$-$^{3}\text{He}$, the potential depths, $V_d$, are found to be 71.21 MeV and 59.26 MeV, with equilibrium distances, $r_d$, of 0.94 fm and 1.28 fm, respectively. The phase shifts increase consistently but nonlinearly. This behavior suggests that the potential has both attractive and repulsive components.

    \item For the $7/2^{-}$ state of $\alpha$-$^{3}\text{H}$ and $\alpha$-$^{3}\text{He}$, the potential depths are 108.16 MeV and 93.06 MeV, with equilibrium distances of 0.88 fm and 1.04 fm, respectively. The scattering phase shifts display a unique trend: they gradually increase up to an energy of approximately 4 MeV, then rise more sharply, leveling off beyond around 6 MeV. This variable phase shift behavior across energy ranges indicates an interplay of attractive and repulsive forces in the potential and highlights resonance effects.
\end{enumerate}
\subsection*{Interaction Potential with centrifugal term:}
After incorporating the centrifugal term into the constructed potentials, a significant reduction in the potential depths was observed for the \(5/2^-\) and \(7/2^-\) states. The modified potential depths, denoted as $\mathbf{V_{d}}^{\text{cent}}$, and their corresponding equilibrium distances, $\mathbf{r_{d}}^{\text{cent}}$, are detailed below:

\begin{itemize}
    \item \textbf{For the \(\alpha\)-\({}^{3}\text{H}\) system:}
    \begin{itemize}
        \item The potential depths reduced to $5.63$ MeV for the \(5/2^-\) state and $15.52$ MeV for the \(7/2^-\) state.
        \item The equilibrium distances shifted to $2.13$ fm for the \(5/2^-\) state and $1.63$ fm for the \(7/2^-\) state.
    \end{itemize}
    
    \item \textbf{For the \(\alpha\)-\({}^{3}\text{He}\) system:}
    \begin{itemize}
        \item The potential depths decreased to $5.42$ MeV for the \(5/2^-\) state and $15.06$ MeV for the \(7/2^-\) state.
        \item The equilibrium distances adjusted to $2.09$ fm for the \(5/2^-\) state and $1.71$ fm for the \(7/2^-\) state.
    \end{itemize}
\end{itemize}

The observed reduction in potential depths and corresponding shifts in equilibrium distances underscore the significant influence of the centrifugal term on the interaction potentials. This adjustment highlights the role of the repulsive centrifugal barrier in modifying the effective interaction between the \(\alpha\)-particle and the target nucleus.

\begin{figure}[htbp]
\begin{tabular}{ccc}
\includegraphics[scale=0.462]{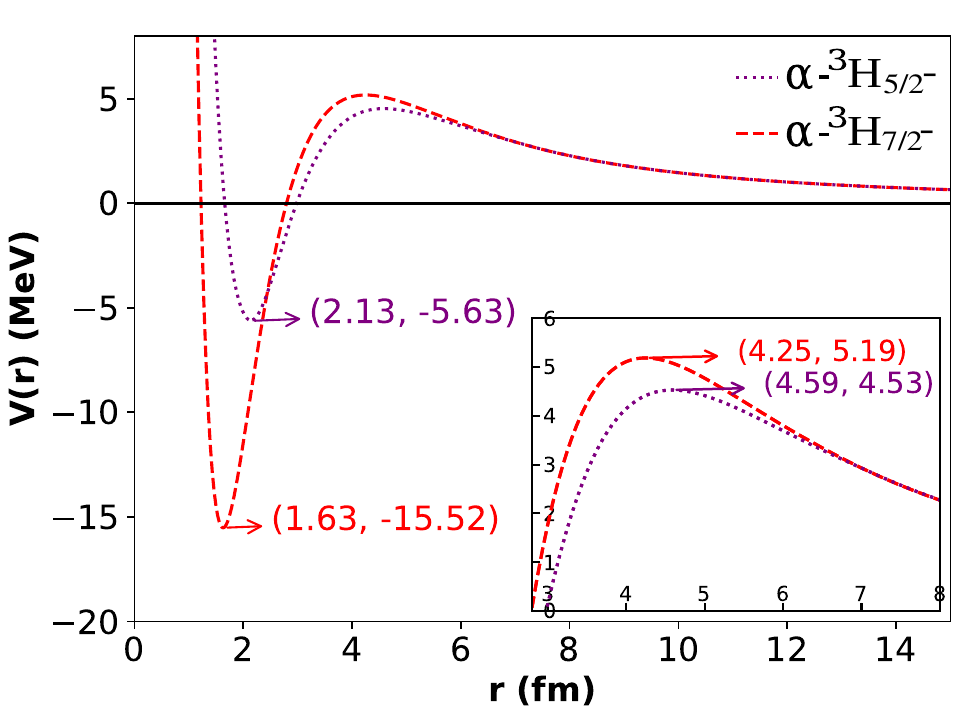}&
\includegraphics[scale=0.462]{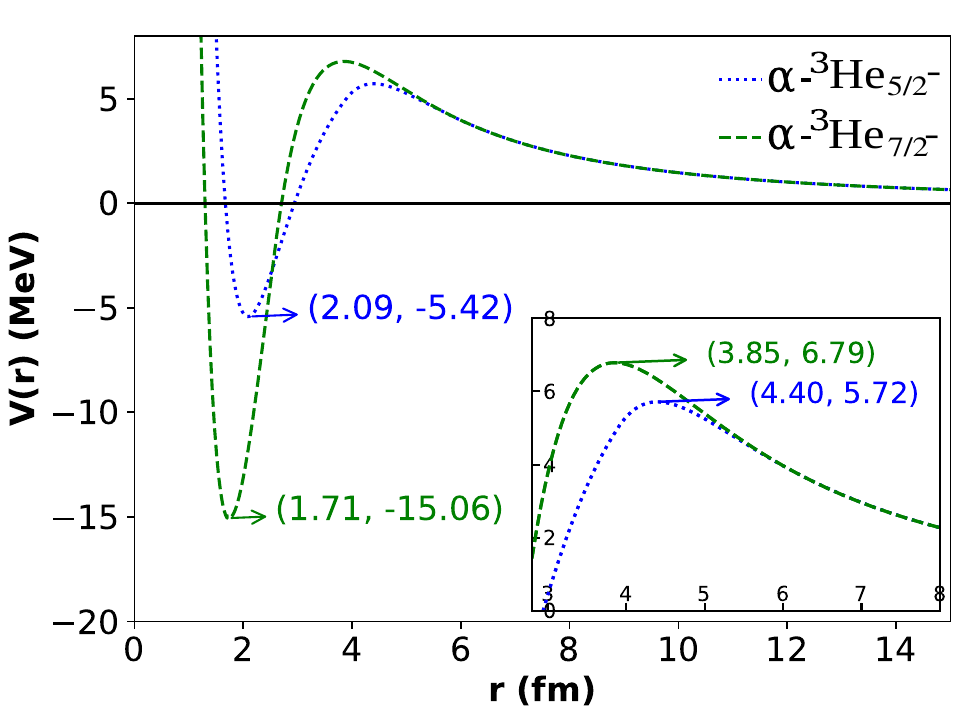}& 
\end{tabular}
 \caption{Interaction potential with centrifugal term for $\alpha$-$^{3}H$ and $\alpha$-$^{3}He$ resonant states, with their coulomb barrier heights (\(V_{CB}\)). }
\label{ptn_ps_cenfu}
\end{figure}

These changes indicate that the centrifugal term alters the spatial characteristics of the interaction, resulting in a shallower potential well and a shift in equilibrium distances. Figure~\ref{ptn_ps_cenfu} shows the modified scattering potentials, including the centrifugal contribution, highlighting the key changes across the interaction range.
This figure further reveals the Coulomb barrier heights (\(V_{CB}\)) and corresponding distances for these interactions. For the \(\alpha\)-\({}^{3}\text{H}\) interaction, the barrier heights are 4.53 MeV and 5.19 MeV, occurring at corresponding distances (\(r_{CB}\)) of 4.59~fm and 4.25~fm for the \(5/2^-\) and \(7/2^-\) states, respectively. Similarly, for the \(\alpha\)-\({}^{3}\text{He}\) system, the barrier heights are 5.72 MeV and 6.79 MeV, occurring at 4.40~fm and 3.85~fm for the \(5/2^-\) and \(7/2^-\) states, respectively. These findings highlight the intricate balance between attractive and repulsive forces in shaping the interaction potential. The modifications introduced by the centrifugal and Coulomb terms are essential for accurately predicting resonance properties and reaction dynamics. This deeper understanding provides a foundation for exploring more complex nuclear systems.

\subsection{Cross section and resonance energy:}    
The partial cross sections $\sigma_\ell(E)$ are calculated using the phase shifts $\delta_\ell(E)$ for each $\ell$, according to the following formula:

\begin{eqnarray*}
    \sigma_\ell(E) = \frac{4\pi(2\ell + 1)}{k^2} \sin^2 \delta_\ell(E)
\end{eqnarray*}

where $k = \sqrt{\frac{2\mu E}{\hbar^2}}$, $\mu$ is the reduced mass, and $\delta_\ell$ denotes the phase shifts for partial waves.

\begin{figure}[htbp]
\begin{tabular}{ccc}
\includegraphics[scale=0.37]{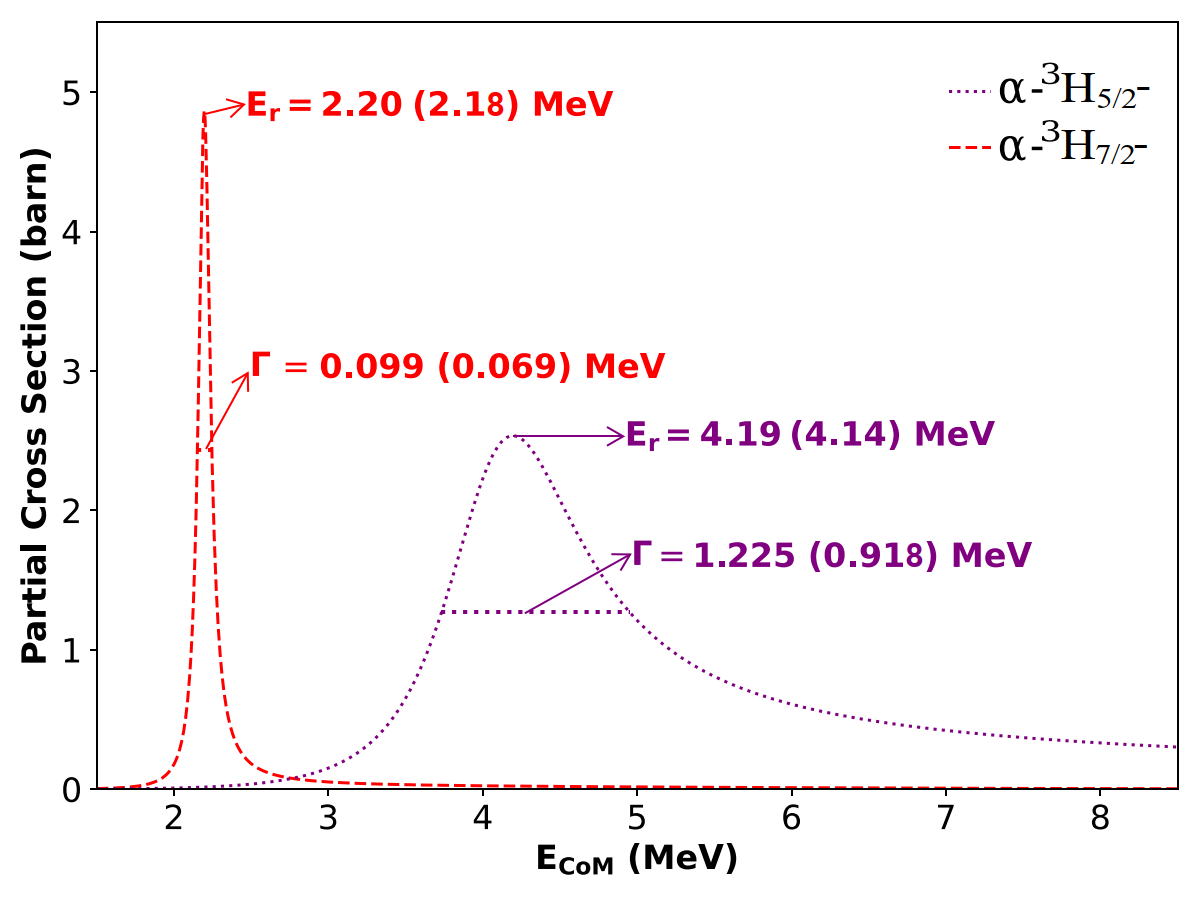}&
\includegraphics[scale=0.37]{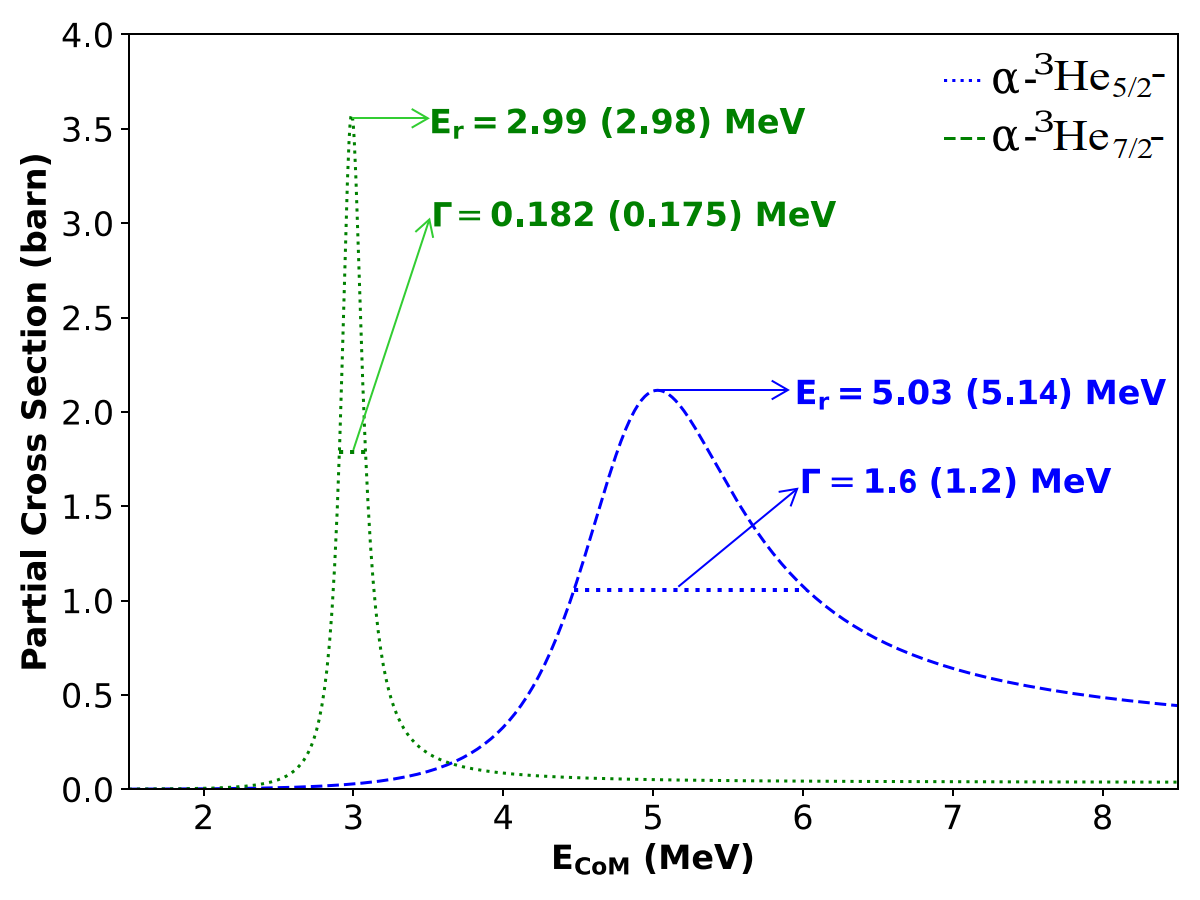}& 
\end{tabular}
 \caption{Corresponding partial cross-section for $5/2^-$ and $7/2^-$ resonating states of $\alpha$-$^{3}H$ and $\alpha$-$^{3}He$ scattering with resonance energy (E$_r$) and full width half maxima ($\Gamma$).}
 \label{cross_section}
\end{figure}

The partial cross sections for various states as a function of $E_{CoM}$, calculated from the obtained phase shift values, are presented in Fig.~\ref{cross_section}. Using these cross sections, we determined the resonance energies (\(E_{r}\)) and decay widths (\(\Gamma\)) for the \(5/2^{-}\) and \(7/2^{-}\) states of both \(\alpha\)-\({}^{3}\text{H}\) and \(\alpha\)-\({}^{3}\text{He}\).

Resonance energies $E_r$ and decay widths $\Gamma$ for the $\alpha$-$^3\text{H}$ and $\alpha$-$^3\text{He}$ systems were calculated and compared with experimental data. For the $\alpha$-$^3\text{H}$ system, a $5/2^-$ resonance was found at an energy $E_r$ of 4.19 MeV with a width $\Gamma$ of 1.225 MeV, while a $7/2^-$ resonance was located at an energy $E_r$ of 5.03 MeV with a width $\Gamma$ of 1.6 MeV. Similarly, for the $\alpha$-$^3\text{He}$ system, a $5/2^-$ resonance was observed at an energy $E_r$ of 2.20 MeV with a width $\Gamma$ of 0.099 MeV, and a $7/2^-$ resonance was identified at an energy $E_r$ of 2.99 MeV with a width $\Gamma$ of 0.182 MeV. These theoretical predictions are in strong agreement with the experimental values reported in Ref. \cite{Dohet}: for the \(\alpha\)-\({}^3\text{H}\) system, resonance energies of 4.14~MeV and 5.14~MeV with corresponding decay widths of 0.918~MeV and 1.2~MeV, respectively, and for the \(\alpha\)-\({}^3\text{He}\) system, resonance energies of 2.18~MeV and 2.98~MeV with corresponding decay widths of 0.069~MeV and 0.175~MeV Ref. \cite{Dohet}. A comparison of the obtained and experimental values is provided in Table~\ref{resonance}. The close agreement between theoretical predictions and experimental values validates the accuracy of the constructed inverse potential and demonstrates its capability to reliably model nuclear interaction dynamics and predict key resonance properties.

\begin{table}[ht]
\centering

\caption{The resonance energies ($E_r$) with resonance width ($\Gamma$) of the $\alpha$-$^{3}$H and $\alpha$-$^{3}$He resonant $f$-wave are presented, with experimental values adopted from \cite{Dohet,Vorabbi}.}
\renewcommand{\arraystretch}{1.2}
\begin{tabular}{cccccc}
\hline
\Xhline{1pt}
      \textbf{System}&\textbf{States}& Exp.\textbf{$E_r$}(MeV) &Sim.\textbf{$E_r$}(MeV) & Exp.\textbf{$\Gamma$}(MeV) & Sim.\textbf{$\Gamma$}(MeV) \\
\hline
\multirow{2}{*}{\textbf{$\alpha$-$^{3}$H}} & ${5/2^-}$ & 4.14 & 4.19 & 0.918 & 1.225 \\  
 & ${7/2^-}$ & 2.18 & 2.20 & 0.069 & 0.099 \\
\hline
\multirow{2}{*}{\textbf{$\alpha$-$^{3}$He}} & ${5/2^-}$ & 5.14 & 5.03 & 1.2 & 1.6 \\  
 & ${7/2^-}$ & 2.98 & 2.99 & 0.175 & 0.182 \\
\hline
\Xhline{1pt}
\end{tabular}

\label{resonance}

\end{table}
\section{Conclusions}
In this research, we have approached the study of nuclear scattering using a machine learning methodology. That is, we obtain the interaction model directly from the available data. Even though we have not modeled the interaction potential using physical considerations, an ab initio functional form consisting of three smoothly joined Morse functions is chosen as the input potential to solve the phase equation. The methodology developed in this work successfully addresses the challenge of incorporating both long-range Coulomb and short-range nuclear interactions in constructing accurate scattering potentials. Our approach effectively accounts for the Coulomb barrier without the need to separately consider the Coulomb potential. Thus, it is possible to utilize the phase function method as an efficient framework to determine phase shifts directly from the potential. Furthermore, the bounds of various parameters were adjusted to ensure that the optimized inverse potentials are physically acceptable. The resonance energies and decay widths for the $\alpha-^{3}H$ and $\alpha-^{3}He$ systems have been obtained and are observed to be in excellent agreement with the experimental data. One can conclude that the ab initio input potential, consisting of three smoothly joined Morse functions, to solve the phase equation of charged particle scattering, such as $\alpha-^{3}H$, $\alpha-^{3}He$, has been successful in the construction of inverse potentials that correctly predict the experimentally observed resonances.

\section*{Acknowledgments}
A. Awasthi acknowledges financial support provided by Department of Science and Technology
(DST), Government of India vide Grant No. DST/INSPIRE Fellowship/2020/IF200538. 
\\ 
\textbf{Author Declaration} 
The authors declare that they have no conflict of interest.
\newpage
\section{Appendix}
\begin{table}[ht!]
\centering
\caption{Expected and simulated scattering phase shifts for $5/2^-$ and $7/2^-$ resonating states of $\alpha$-$^{3}H$ system for lab energies below 11 MeV are shown in Table below. The final datasets used for calculation are highlighted in the Table.}
\label{table1}
\scalebox{1}{
\begin{tabular}{c c c c c c}
\hline 
\Xhline{1pt}
$E_{lab}(MeV)$ &~ $\delta_{5/2^-}^{exp}$\cite{Spiger} & ~$\delta_{5/2^-}^{3Morse}$ & $E_{lab}$(MeV) & ~$\delta_{7/2^-}^{exp}$\cite{Spiger} & ~$\delta_{7/2^-}^{3Morse}$\\
\hline
5.22  & 0.127  & 3.428  &\textbf{ 3.88}  &\textbf{ 2.195}  &\textbf{ 2.429}\\
5.34  & 0.719  & 3.746  & 4.02  & 4.247  & 3.019\\
5.53  & 1.903  & 4.302  & 4.16  & 6.557  & 3.797\\
5.74  & 3.679  & 5.001  &\textbf{ 4.43}  & \textbf{8.113}  & \textbf{6.253}\\
\textbf{6.02}  & \textbf{4.863}  & \textbf{6.093}  & 4.46  & 11.175  & 6.654
\\
\textbf{6.26  }&\textbf{ 6.342 } & \textbf{7.203}  & 4.74  & 14.693 & 13.472\\
\textbf{6.53}  & \textbf{7.822}  & \textbf{8.684}  &\textbf{ 4.82 } & \textbf{20.543} & 17.665\\
\textbf{6.77}  & \textbf{10.486} & \textbf{10.253} & 5.0   & 37.733 & 42.219\\
\textbf{7.04}  & \textbf{11.670}& \textbf{12.369} & \textbf{5.03}  & \textbf{49.248 }& \textbf{51.439}\\
\textbf{7.29}  & \textbf{14.926} & \textbf{14.741} & 5.04  & 68.907 & 55.102\\
\textbf{7.54}  & \textbf{17.590 }& \textbf{17.612} & \textbf{5.07}  &\textbf{ 69.075} & \textbf{67.973 }\\
\textbf{7.81}  & \textbf{21.734} &\textbf{ 21.426} & \textbf{5.18}  & \textbf{121.455} & \textbf{120.886 }\\
\textbf{8.06}  & \textbf{26.765}  & \textbf{25.795}  & 5.28  & 137.608 & 144.858\\
\textbf{8.32}  & \textbf{30.909} & \textbf{31.417} &\textbf{ 5.3}   &\textbf{ 146.905 }& \textbf{147.601}\\
\textbf{8.53}  & \textbf{38.013} & \textbf{36.929} & 5.55  & 165.320 & 163.073\\
\textbf{8.79}  & \textbf{46.300}& \textbf{45.135} & \textbf{5.66}  & \textbf{167.62} & \textbf{165.691}\\
\textbf{9.06}  & \textbf{55.180}&\textbf{ 55.320} &\textbf{ 6.53}  & \textbf{172.651} & \textbf{172.154}\\
\textbf{9.3}   & \textbf{64.651} & \textbf{65.541} & 7.27  & 176.289 & 173.457\\
\textbf{9.58}  &\textbf{ 76.490}& \textbf{78.061} &\textbf{ 8.04 } & \textbf{175.018} & \textbf{174.058 }\\
\textbf{9.82}  & \textbf{88.330}& \textbf{88.436} & 8.27  & 176.289 & 174.174\\
\textbf{10.06} & \textbf{99.281} & \textbf{97.815} &\textbf{ 9.06}  & \textbf{176.794} & \textbf{174.466}\\
10.32 & 110.825 & 106.502 & -- & -- & -- \\
\Xhline{1pt}
\hline
\end{tabular}
}
\end{table}
\begin{table}[ht!]
\centering
\caption{Expected and simulated scattering phase shifts for $5/2^-$ and $7/2^-$ resonating states of $\alpha$-$^{3}He$ system for lab energies below 11 MeV are shown in Table below. The final datasets used for calculation are highlighted in the Table.}
\label{table2}
\begin{tabular}{cccccc}
\hline 
\Xhline{1pt}
$E_{lab}$(MeV) &~ $\delta_{5/2^-}^{exp}\cite{Boykin,Hardy}$ & ~$\delta_{5/2^-}^{3Morse}$ & $E_{lab}(MeV)$ & ~$\delta_{7/2^-}^{exp}\cite{Boykin,Hardy}$ & ~$\delta_{7/2^-}^{3Morse}$ \\
\hline
\textbf{3.3}& \textbf{2}& \textbf{1}& \textbf{3.3} & \textbf{2} & \textbf{1.2} \\
\textbf{5.21}& \textbf{3}& \textbf{4.9}& 3.51 & \textbf{2} & \textbf{1.7} \\
\textbf{5.69}& \textbf{8}& \textbf{7.1}& \textbf{3.88} &\textbf{ 3} & \textbf{2.8} \\
6.45& 11.9& 13.0& \textbf{4.37} &\textbf{7} & \textbf{6.1} \\
\textbf{7.2}& \textbf{23.9}& \textbf{24.0}& 4.64 & 10 & 10.3 \\
\textbf{8.2}& \textbf{56.7}& \textbf{56.3}& \textbf{4.79} & \textbf{15} & \textbf{14.8} \\
8.46& 68.2& 68.7& \textbf{5.21} &\textbf{ 78} & \textbf{78.1} \\
\textbf{8.71}& \textbf{80.7}& \textbf{80.8}& \textbf{5.69} & \textbf{157.4} & \textbf{157.1} \\
8.96& 89.2& 92.1&\textbf{ 6.19} & \textbf{166.2} & \textbf{166.1} \\
\textbf{9.21}& \textbf{101}& \textbf{101.8}&\textbf{ 6.45} & \textbf{167.6} & \textbf{167.8} \\
\textbf{9.71}& \textbf{116.8}& \textbf{116.0}& \textbf{6.95} &\textbf{ 169.6} & \textbf{169.6} \\
9.96& 119.4& 121.1& 7.2 & 170.9 & 170.1 \\
\textbf{10.96}& \textbf{132.8}& \textbf{133.1}& 7.7 & 171 & 170.7 \\
-- & -- & -- & \textbf{7.95} & \textbf{172} & \textbf{170.9} \\
-- & -- & -- & 8.2 & 173 & 171.0 \\
\hline
\Xhline{1pt}
\end{tabular}
\end{table}

\bibliographystyle{unsrt}  

\end{document}